\documentclass[a4paper,twoside,openright,12pt,prb]{article}
\usepackage{graphics}

\begin{document}

\title{Static properties of the dissipative random quantum
Ising ferromagnetic chain}
\author{L. F. Cugliandolo,$^{1,2}$ G. S. Lozano$^{3}$ 
and H. Lozza$^{3}$\\
{\footnotesize \it
$\;^1$ Laboratoire de Physique Th{\'e}orique 
et Hautes {\'E}nergies, Jussieu,} \\
{\footnotesize \it
4 Place Jussieu, 75252 Paris Cedex 05, France} \\
{\footnotesize \it
$\;^2$ Laboratoire de Physique Th{\'e}orique, 
{\'E}cole Normale Sup\'erieure de Paris,} \\
{\footnotesize \it 
24 rue Lhomond, 75231 Paris Cedex 05, France}, \\
{\footnotesize \it
$\;^3$ Departamento de F\'{\i}sica,
Universidad de Buenos Aires,} \\
{\footnotesize \it
Pab. I, Ciudad Universitaria 1428, Argentina.}
}
\date{\today}

\maketitle 

\begin{abstract}
We study the zero temperature static properties of dissipative
ensembles of quantum Ising spins arranged on periodic one dimensional
finite clusters and on an infinite chain. The spins interact
ferro-magnetically with nearest-neighbour pure and random
couplings. They are subject to a transverse field and coupled to an
Ohmic bath of quantum harmonic oscillators. We analyze the coupled
system using Monte Carlo simulations of the classical two-dimensional
counterpart model. The coupling to the bath enhances the extent of the
ordered phase, as found in mean-field spin-glasses.  In the case of
finite clusters we show that a generalization of the Caldeira-Leggett
localization transition exists.  In the case of the infinite random
chain we study the effect of dissipation on the transition and the
Griffiths phase.
\end{abstract}
{\tt KEYWORDS: 
Spin chain, disorder, dissipation, Monte Carlo}
\newpage

\section{Introduction}
\label{sec:intro}

 The influence of quenched disorder~\cite{Fisher,Vojta} on the one
hand and the effect of the coupling to an environment~\cite{Troyer,Pankov},
on the other, on {\it finite} dimensional quantum spin models have
been separately analyzed recently.  The former introduce Griffiths-Mc Coy
singularities that are particularly important for quantum
systems~\cite{Fisher,McCoy,Sachdev}.  Dissipation implies decoherence
and the coupling to a quantum bath generates highly non-trivial
localization phenomena at least for a single two-level
system~\cite{Leggett}.  The combined effect of quenched disorder and
the coupling to a quantum reservoir has only been analyzed in
mean-field models so far~\cite{bagno1,bagno2}.
 
A ring of quantum Ising spins is probably the simplest finite
dimensional model exhibiting a phase transition from a paramagnetic
({\sc pm}) to a ferromagnetic ({\sc fm}) phase where one can analyze
the mixed effect of disorder and dissipation.  Indeed, the bath
introduces long-range ferromagnetic interactions in the imaginary-time
direction that add to the nearest-neighbor ferromagnetic couplings
due the quantum nature of the spins and, at zero temperature, one
finds a phase transition even for finite size clusters.  The $d + 1$
equivalent classical problem is characterized by perfect correlation
in the imaginary-time direction and the effect of disorder is
consequently very strong. For sufficiently large system sizes and, in
particular, in the thermodynamic limit rare regions in the sample may
have a very strong effects on the properties close to the phase
transition, giving rise to very interesting Griffiths phenomena.

The zero-temperature critical behavior of the isolated finite
dimensional infinite system in one, two and three dimensions has been
studied in detail in a series of analytic~\cite{Fisher,Igloi,Monthus} 
and numerical
papers~\cite{Rieger-young,Guoetal,Young-Rieger,Pich,Motrunich,Rieger-review}.  
An `infinite
randomness fixed point scenario', in which the system appears more and
more random at larger and larger scales, emerged~\cite{Fisher}.  The
critical behavior is quite uncommon, with activated scaling on the
transition and very large distributions of local properties in a
finite interval close to the quantum critical point that lead to
different typical and average values and the divergence of the global
susceptibility well within the paramagnetic phase.

In this paper we investigate the effect of Ohmic dissipation on the
equilibrium properties of small clusters and an infinite chain of
quantum Ising spins with ferromagnetic exchange interactions of pure
and random type.  We focus on the following questions: (i) How is the
Caldeira-Leggett localization transition modified when several
ferro-magnetically interacting spins are coupled to an environment.
(ii) Which are the characteristics of the quantum phase transition, as
a function of the transverse field and the coupling strength to the
bath, for finite clusters and in the thermodynamic limit.  (iii)
Whether the peculiar phenomenology of the isolated random Ising
chain~\cite{Fisher}, especially in its paramagnetic phase, also exists
when it is coupled to an environment. Conflicting predictions, based
on phenomenological droplet-like arguments~\cite{Antonio} and the
analysis of some simplified models~\cite{Millis}, appeared in the
literature recently.  We address these questions here using Monte
Carlo simulations of the equivalent classical model in $d=2$
with short-range interaction in the spatial direction and long-range
interactions in the imaginary-time direction as introduced by the 
coupling to the bath.

The paper is organized as follows. In Sect.~\ref{sec:themodel}
we define the model. Section~\ref{sec:MC} is devoted to a brief 
description of the Monte Carlo method used to generate the numerical 
data. We also summarize in this Section the parameters studied.
In Sect.~\ref{sec:results} we explain our results. Finally, 
in Sect.~\ref{sec:conclusions} we present our conclusions and
discuss some areas of applicability of our results.

\section{The model}
\label{sec:themodel}

The Hamiltonian of the random quantum Ising spin chain is
\begin{equation}
H_J = - \sum_{i=1}^N J_{i} \hat\sigma^z_i
\hat\sigma^z_{i+1} - \sum_{i=1}^N \Gamma_i \hat\sigma^x_i
 - \sum_{i=1}^N h_i \hat \sigma^z_i
\; .
\end{equation}
The spins lie on the vertices of a periodic one dimensional lattice
with $N$ sites.  They are represented by Pauli matrices satisfying the
SU(2) algebra: $[\hat\sigma_i^a, \hat\sigma_j^b] =\delta_{ij}
\epsilon_{abc} \imath \hat \sigma_i^c$ with $a,b,c=1,2,3$ associated
to $x,y,z$, respectively and $i = 1, \cdots,N$.  We consider $J_i$,
the strength of the exchange interactions between nearest-neighbors,
as independent quenched random variables in the interval $ 0< J_i<1 $
with uniform probability density.  Without loss of generality, we
choose $J_i$ positive since for one dimensional lattices the sign of
the interactions can be gauged away (even when the system is linearly
coupled to the coordinates of the harmonic oscillators representing
the bath).  The next-to-last term is a coupling to a local transverse
field that, for simplicity, we choose to be the same on all sites,
$\Gamma_i=\Gamma$. The last term is a coupling to a longitudinal field
that one may include to compute local susceptibilities.

The chain plus environment is modeled by
\begin{equation}
\hat H = \hat H_J + \hat H_B + \hat H_I + \hat H_{CT}
\end{equation}
where $\hat H_B$ is the Hamiltonian of the bath, $H_I$ 
represents the interaction between the chain and the bath and 
$H_{CT}$ is a counter-term that serves to eliminate an undesired 
mass renormalization induced by the coupling~\cite{Weiss}.  
We assume that each 
spin in the system is coupled to its own
set of $\tilde N/N$ independent harmonic oscillators with $\tilde N$ 
the total number of them.  
The bosonic Hamiltonian for the isolated 
reservoir is 
\begin{equation}
\hat H_B = \sum_{l=1}^{\tilde N} \frac{1}{2m_l} \hat p^2_l 
+ \sum_{l=1}^{\tilde N}  \frac{m_l \omega_l^2}{2}
\hat x^2_l 
\; .
\end{equation} 
The coordinates $\hat x_l$ and the momenta $\hat p_l$ satisfy
canonical commutation relations.  
For simplicity we consider a bilinear coupling, 
\begin{equation}
\hat H_I = -\sum_{i=1}^{N} \hat \sigma_i^z 
\sum_{l=1}^{\tilde N} c_{il} \, \hat x_l
\; ,
\end{equation}
that involves only the oscillator coordinates.  The 
counter term reads 
\begin{equation}
\hat H_{C T} = \sum_{l=1}^{\tilde N}\frac{1}{2m_l \omega_l^2} 
\left( \sum_{i=1}^N
c_{il} \, \hat \sigma_i^z \right)^2  \; .
\end{equation} 

The partition function of the whole system for a particular
realization of the random exchange interactions is $Z_J =
\mbox{Tr} e^{-\beta \hat H}$, involving a sum over all states of the
chain and the bath.  This problem can be mapped onto a classical model
by using the Trotter-Suzuki formalism that amounts to writing the
path-integral for the partition function as a sum over spin and
oscillator variables evaluated on a discrete imaginary-time grid,
$\tau_t = t \Delta \tau$ with $\Delta \tau \equiv \beta\hbar /N_\tau$,
labeled by the index $t = 0, \cdots, N_\tau-1$.  We assume periodic
boundary conditions in all directions with $\beta \hbar$ the length in
the $\tau$ direction. This mapping is exact in the limit $N_\tau \to
\infty$. In what follows, we use units such that $k_B=\hbar=1$.

The integration over the bath variables can be 
performed explicitly; for an Ohmic bath the resulting classical action 
reads 
\begin{eqnarray}
{\cal A}[K_i, B, \alpha;s_i^t] &=&
-\sum_{t=0}^{N_\tau-1}
\sum_{i=1}^N K_i s_i^t s_{i+1}^t
- B
\sum_{t=0}^{N_\tau-1} \sum_{i=1}^{N} s_{i}^t s_{i}^{t+1}
\nonumber \\
& &  -
\frac{\alpha}{2} \sum_{t<t'}^{N_{\tau}-1} \sum_{i=1}^{N}
\left( \frac{\pi}{N_\tau} \right)^2
\frac{s_i^{t} s_i^{t'}}{\sin^2\left( \pi |t-t'|/N_\tau\right)}
\; ,
\label{classicalA}
\end{eqnarray}
where we set $h_i=0$ and we dropped an irrelevant constant factor.
$s_i^t$ are $N\times N_\tau$ classical Ising variables, $s_i^t=\pm 1$,
representing the $z$ component of the quantum spin at each instant
$\tau_t$ at site $i$.  Notice that the bath is responsible for the
appearance of long range ferromagnetic interactions in the $\tau$
direction controlled by the chain-bath coupling parameter $\alpha$.
From the quantum nature of the spins, a nearest-neighbour
ferromagnetic coupling also arises in the $\tau$ direction. Its
strength $B$ is defined by
\begin{equation}
\exp{-2 B}= \tanh(\Delta\tau\Gamma) \; ,
\label{defB}
\end{equation}
and describes the intensity of the quantum fluctuations.  Further, since
$J_i$ are quenched random variables, the random values of the spatial
interactions
\begin{equation} 
K_i=\Delta \tau J_i
\label{defK}
\end{equation}
are the same along each imaginary-time direction, {\it i.e.} depend
only on the spatial label $i$.

\section{Monte Carlo simulations}
\label{sec:MC}

We analyze the static properties of the system by means of Monte Carlo
simulations applied to the effective $(1 + 1)$ dimensional partition
function,
\begin{equation}
Z_J = \sum_{s_i^t=\pm 1} e^{-{\cal A}[K_i,B,\alpha;s_i^t]} \; .
\end{equation}
The sum over all spin configurations accounts for the statistical 
average that we denote, henceforth, with angular brackets. In the 
disordered cases, the thermodynamic properties and, in particular, 
the phase transition follow from the analysis of the free-energy 
density averaged over the probability distribution of the 
random exchanges. We indicate this average with square brackets. 

Taking advantage of the fact that the ferromagnetic models are not
frustrated (all $J_i > 0)$, we use an efficient Monte Carlo algorithm
that adapts the cluster updates~\cite{Wolff} to the case with long
range ferromagnetic interactions~\cite{luijten}.  In addition, in
order to avoid running several simulations with similar parameters, we
use histogram methods to scan the space of parameters near a phase
transition~\cite{Ferrenberg}.

The classical counterpart model is defined on a rectangular lattice
with size $N\times N_\tau$. The zero temperature limit,
$\beta\to\infty$, is achieved by taking the thermodynamic limit in the
imaginary time direction, $N_\tau\to\infty$.  We used various system
sizes varying the aspect ratio of the $2d$ sample. We took $N$ form 1
to 32 and $N_\tau$ form 4 to 1024.  We found that up to $400$ Monte
Carlo sweeps were needed to equilibrate the largest sample sizes (note
the efficiency of the cluster algorithm!).  We simulated up to 2048
different realizations of the exchange interactions to account for the
effect of disorder.
 
We concentrate on the critical properties of finite and infinite
periodic chains at zero temperature.  As we have already set the
disorder scale $0<J_i<1$, a complete description of the state of the
system is given by the dimensionless parameters $\alpha$ and $\Delta
\tau \Gamma $.  For comparison, we also simulate the non-random chain
coupled to an Ohmic bath replacing the random variables $J_i$ with
uniform distribution between $0$ and $1$ by a constant value equal to
the maximum, $J_i=1$.

\section{Results} 
\label{sec:results}

In this Section we present the results of our numerical 
simulations. 

As mentioned before, we used system sizes ranging form $N=1$ to
$N=32$. We separate our results in two groups, small $N$ ($N\leq 4$),
and large $N$ ($N>4$). In the first group we intend to study whether
the random exchange interactions modify the physics of the well known
Caldeira-Leggett model while for the second group our results
generalize those reported in~\cite{Troyer} (where no random
interactions are present) and those reported
in~\cite{Fisher,Young-Rieger,Pich,Crisanti} (where no bath is
present).

\subsection{Phase diagram}

\subsubsection{Analysis of the Binder parameter}

The phase transition in the $(\Delta \tau \Gamma,\alpha)$-plane 
can be found by analyzing the Binder ratio 
\begin{equation}
g_{av} 
\equiv \frac12 \left[ 3 - \frac{\langle m^4\rangle}
{\langle m^2\rangle^2}\right]
\; ,
\label{defbinder}
\end{equation}
where the angular brackets indicate the statistical average and the
square brackets denote the average over disorder (if there is no
disorder, this step is not necessary).  The magnetization density is defined
as 
\begin{equation}
m\equiv \frac{1}{N N_\tau} 
\sum_{i=1}^N \sum_{t=0}^{N_\tau-1} s_{i}^{t}
\; . 
\end{equation}
The analysis of the Binder ratio presents different aspects whether we
consider a finite cluster or its thermodynamic limit.

\vspace{0.25cm}
\noindent\underline{\emph{(i) Small $N$}}
\vspace{0.25cm}

When $N$ is finite the zero-temperature limit of the quantum 
problem corresponds to a classical model defined on 
a stripe of finite width ($N<\infty$) and infinite height
$(N_\tau\to\infty)$. In this case, the usual finite size 
scaling analysis of the adimentional Binder parameter can be applied.
The critical values are easily found by plotting $g_{av}$ as a
function of $\alpha$ for several values of $N_\tau$.  The
dimensionless combination of momenta in Eq.~(\ref{defbinder}) yields
$g_{av}$ independent of size at criticality. Then, $\alpha_c$ can be
located by finding the intersection of the $g_{av}$ curves against
$\alpha$ for several $N_\tau$~\cite{librobinder} for every choice 
of $\Delta\tau_c\Gamma_c$.  

\begin{figure}[t]
\resizebox{13cm}{!}{\includegraphics*[1.5in,4.7in][7in,7.7in]{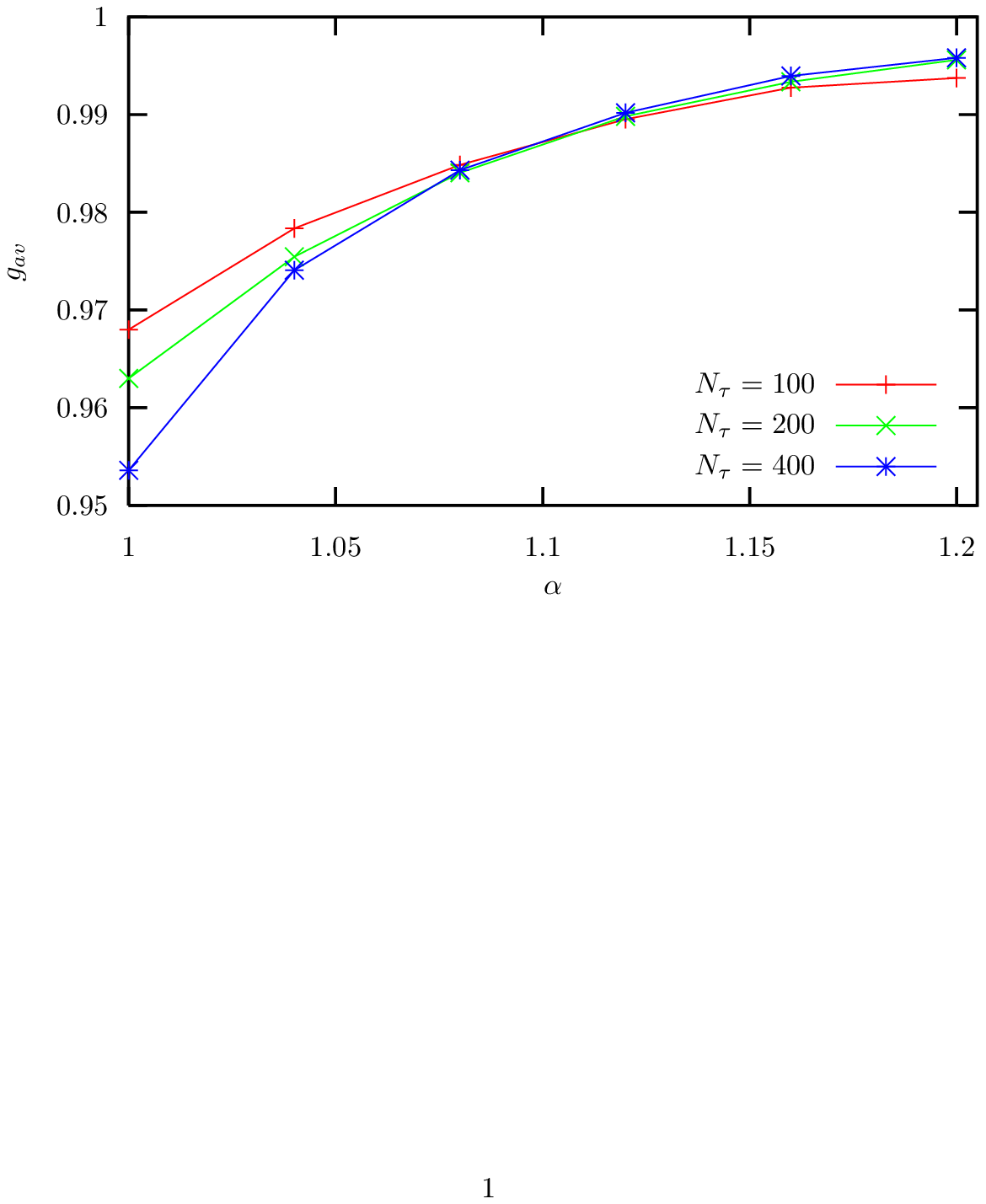}}
\caption{(Color on-line.) The Binder cumulant as a function of $\alpha$
for $N=1$ with $\Delta \tau_c \Gamma_c=0.48$. The localization
transition occurs at $\alpha_c=1.10 \pm 0.02$.}
\label{binderN1}
\end{figure}

The case $N=1$ corresponds to the well known problem of a two level
system coupled to a bath. In this case we expect a transition from an
incoherent state to a localized one at $\alpha_c=1.0$ as analytically
demonstrated by Leggett {\it et al.}~\cite{Leggett}. As a test for our
numerical algorithm, we re-derive this result numerically.
In Fig.~\ref{binderN1} we show an example with $N=1$ and
$\Delta \tau_c \Gamma_c=0.48$, where we see that the curves cross at
$\alpha_c=1.10 \pm 0.02$ (the error measures the precision on the
crossing point). We build the phase
diagram by repeating these steps for several values of $\Delta \tau_c
\Gamma_c$.

\begin{figure}[t]
\resizebox{13cm}{!}{\includegraphics*[1.5in,4.7in][7in,7.7in]{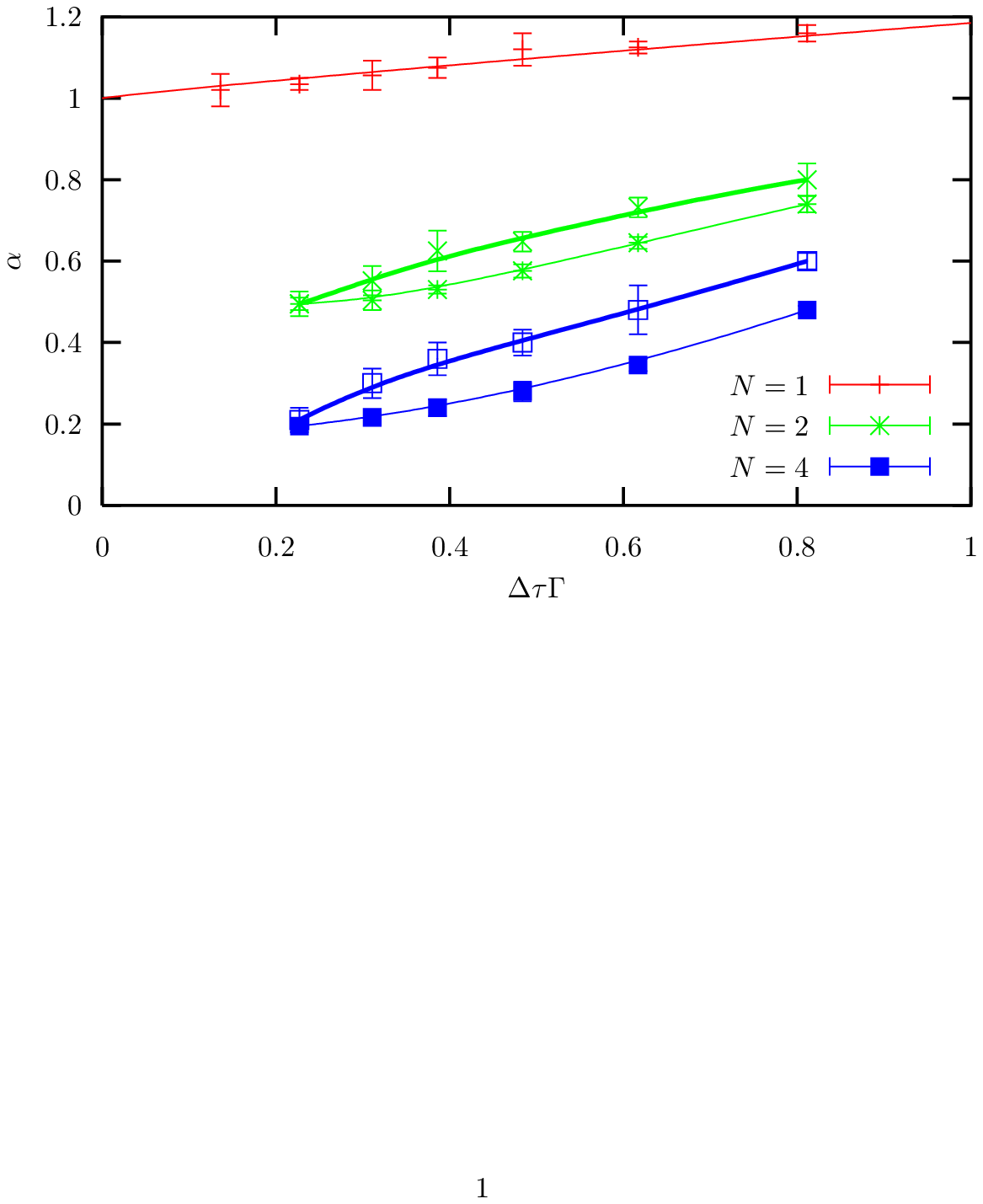}}
\caption{(Color on-line.) Critical boundary 
$(\Delta\tau_c\Gamma_c,\alpha_c)$ 
for small clusters, $N\leq4$. For $ N=1$
this corresponds to the localization transition in the
Caldeira-Leggett model. The data are well fitted by a linear function
reaching $\alpha_c=1$ at $\Delta \tau_c \Gamma_c=0$. For $N>1$, we
show the transition curve for systems with random exchange
interactions (thick lines) and non-random exchange interactions (thin
lines).}
\label{dfsmallN}
\end{figure}

We use the same method to obtain the critical line for $N=2$ and $4$ both
for random and non-random exchange interactions.  In
Fig.~\ref{dfsmallN} we show the phase diagram for $N=1,2,4$. In
general, it is hard to determine $\alpha_c$ for
$\Delta \tau \Gamma < 0.2$ where $g_{av}$ varies less than
0.1\% and $\alpha_c$ is masked by the noise.

For $N=1$ the critical values $(\Delta\tau_c\Gamma_c,\alpha_c)$
are well fitted by a linear function that for $ \Delta
\tau_c \Gamma_c \to 0$ approaches the analytical result
$\alpha_c=1$~\cite{Leggett}.  The localization
transition moves to lower values of $\alpha_c$ when $N$ increases. 
In addition, for finite $N$, the disordered chain
has a lower $\Delta \tau_c \Gamma_c$ for a given $\alpha_c$.  However,
this difference tends to reduce when $ \Delta \tau_c \Gamma_c \to 0$.

\newpage

\vspace{0.25cm}
\noindent\underline{\emph{(ii) Thermodynamic limit}}
\vspace{0.25cm}

In the thermodynamic limit the finite size scaling analysis 
must be done using the parameters $N$ and $N_\tau$ and, 
naively, one is forced to assume a scaling relation between 
them. The Binder cumulant is an adimentional quantity that
on the phase transition must scale as
\begin{eqnarray}
g_{av} &\sim& 
\tilde g(N/\xi, y)
\label{binder-scaling}
\end{eqnarray}
where $\xi$ is the spatial correlation length, 
$\tilde g$ is a scaling function and $y$ is a ratio between the 
imaginary-time size $N_\tau$ and some adequate function of the 
spatial size $N$. For instance, one has 
\begin{eqnarray}
y = \left\{
\begin{array}{ll}
N_\tau/N^z \;\; 
&
\mbox{conventional scaling} \;, 
\\
N_\tau/e^{N^{\overline z}} 
\;\;
&
\mbox{activated scaling} 
\; .
\end{array}
\right.
\end{eqnarray}

\begin{figure}[p]
\resizebox{10cm}{!}{\includegraphics*[1.5in,4in][7in,7.3in]{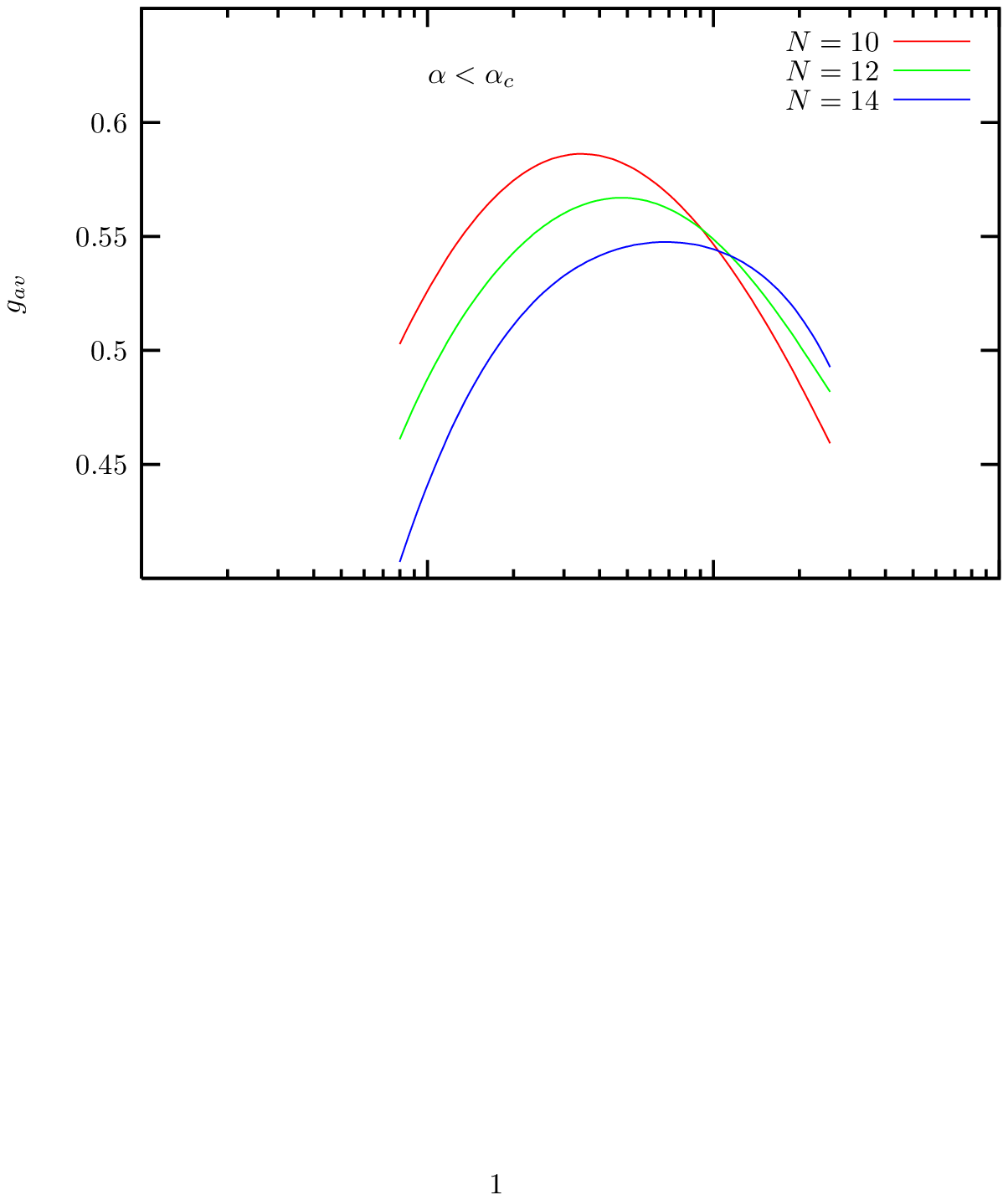}}
\resizebox{10cm}{!}{\includegraphics*[1.5in,4in][7in,7.3in]{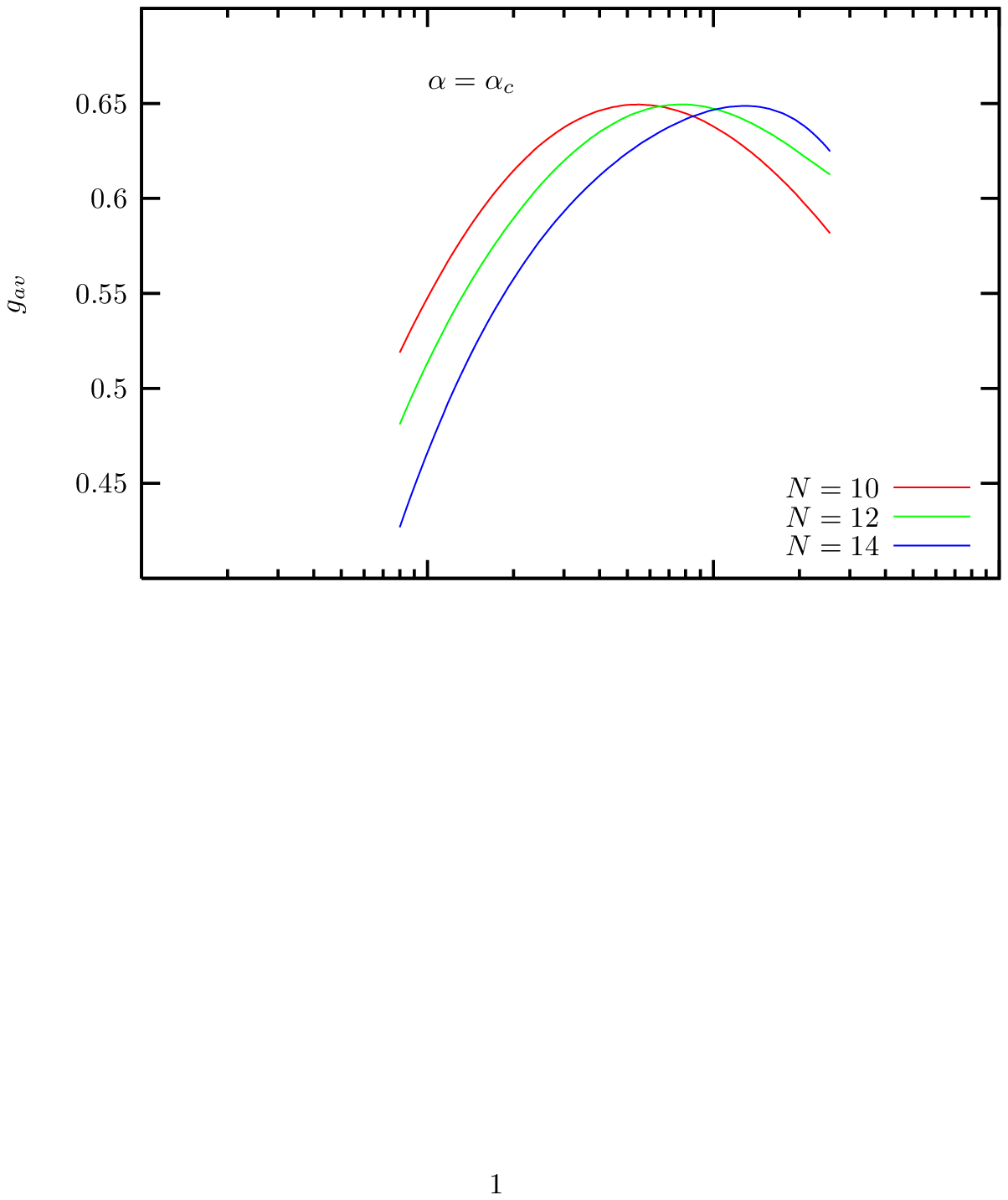}}
\resizebox{10cm}{!}{\includegraphics*[1.5in,4in][7in,7.3in]{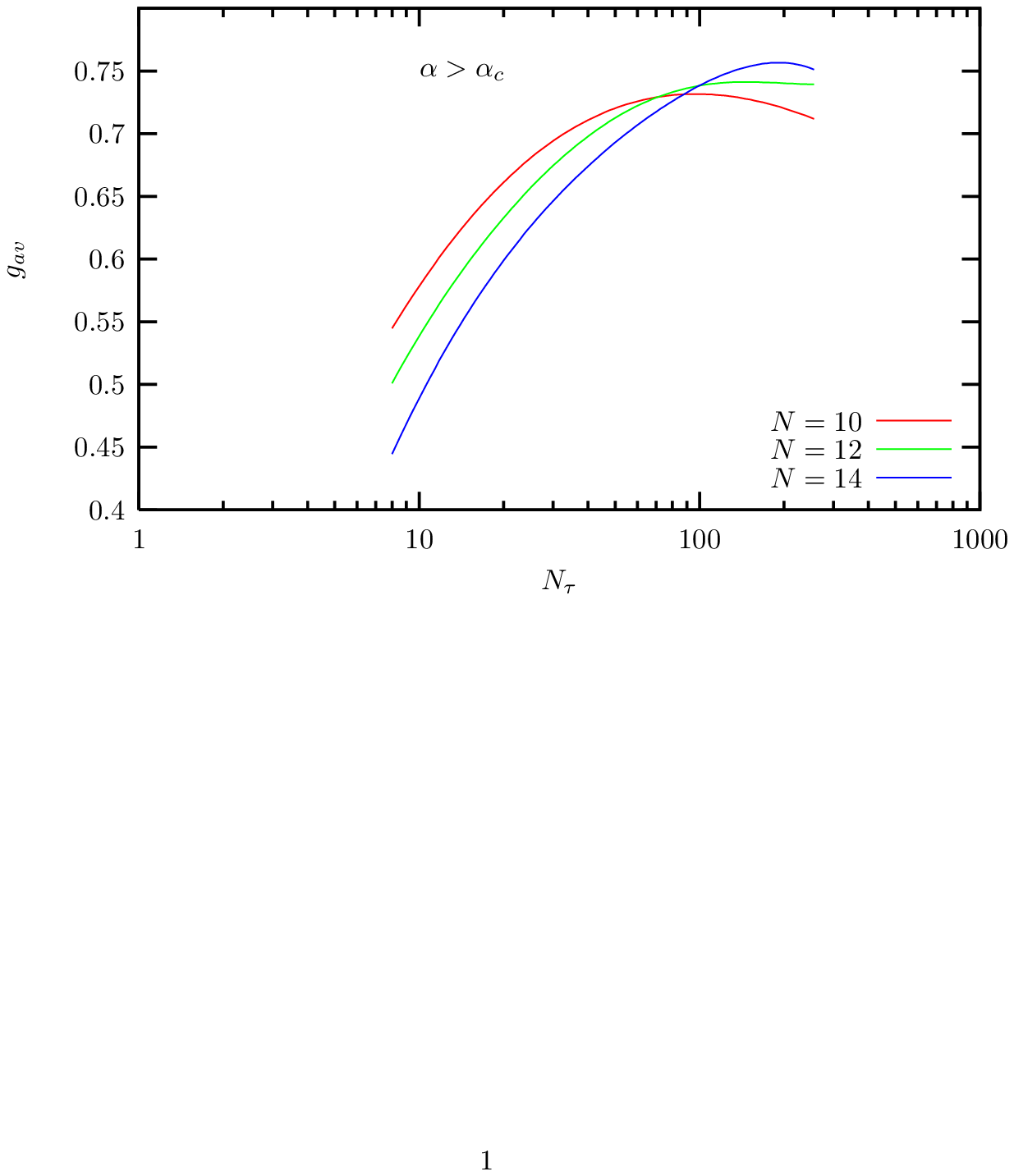}}
\caption{(Color on-line.) Typical behavior of the Binder cumulant at fixed
$\Delta\tau_c\Gamma_c$ and in a neighborhood of its corresponding
$\alpha_c$. The maximum of
$g_{av}$ is independent of $N$ only at $\alpha=\alpha_c$, see the
central panel.}
\label{bindmim}
\end{figure}

\begin{figure}[t]
\begin{center}
\resizebox{10cm}{!}{\includegraphics*[1.5in,4.in][7in,7.7in]{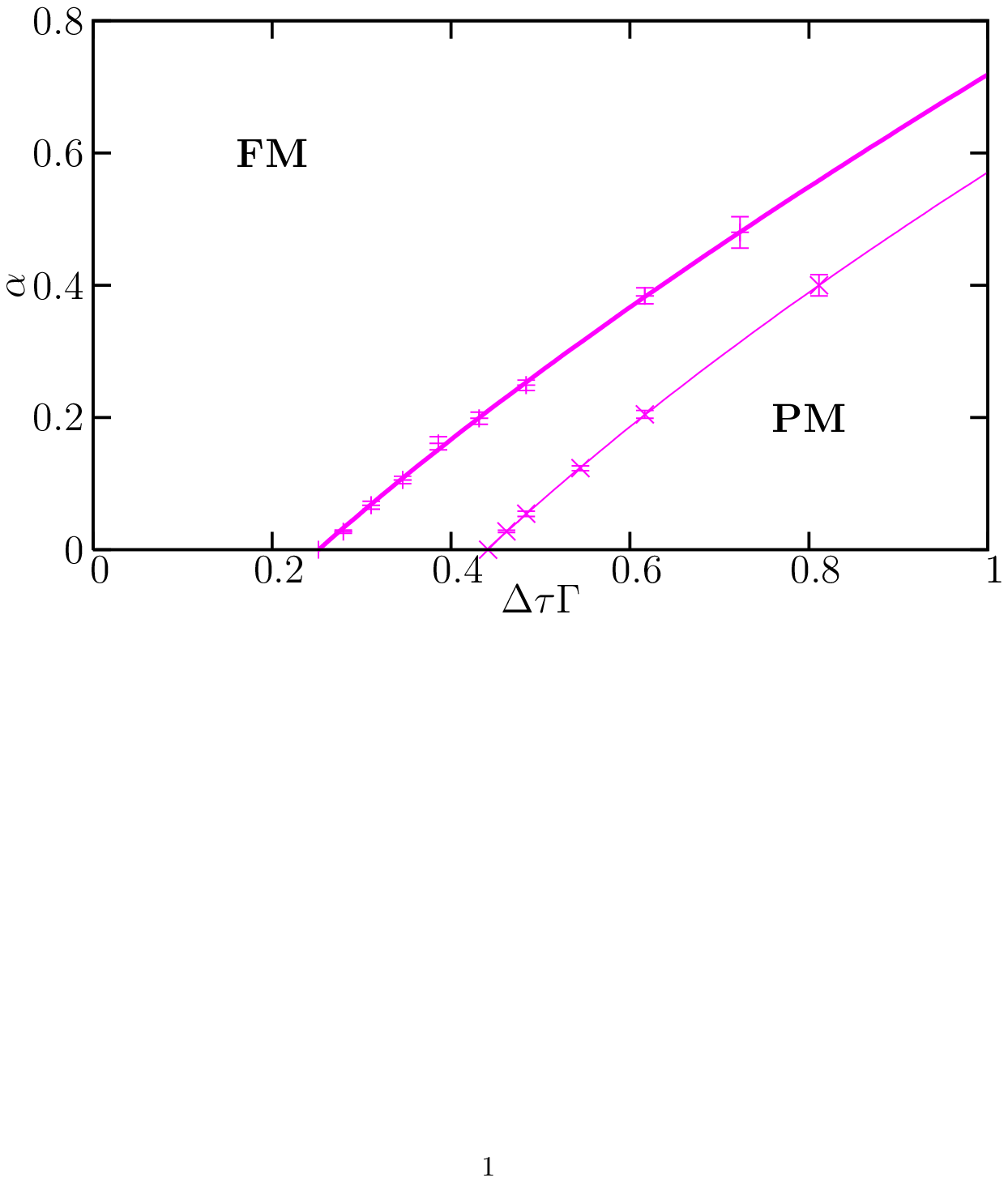}}
\end{center}
\caption{(Color on-line.)  Phase diagram for $N=\infty$.  We show this
line for the disordered chain with a uniform distribution of exchanges 
on the interval $[0,1]$ (thick lines) and the non-random chain
with $J=1$ (thin lines).}
\label{thermolimit}
\end{figure}

A simple argument shows that for fixed $N$ and generic values of the
other parameters the Binder cumulant attains a maximum as a function of
$N_\tau$~\cite{Rieger-young,Guoetal}.  This argument still holds when
long-range interaction are introduced by the bath.  Indeed, when
$N_\tau$ is very small with respect to $N$, one effectively has a very
long one-dimensional stripe (along the spatial dimension) and the
long-range interactions induced by the bath on the imaginary-time
direction are irrelevant. Moreover, for the values of the parameters
that are close to the phase transition of the infinite $2d$ system,
one is well above the transition of the one dimensional system and
$g_{av}\to 0$. In the opposite limit of $N_\tau \gg N$ one is back in
the kind of finite cluster problem discussed in the previous
paragraphs.  Now, for the parameters chosen, the new one-dimensional
model should have a finite correlation length away from its critical
line and $g_{av}\to 0$ as well.  For $N$ and all other parameters
fixed $g_{av}$ should then reach a maximum as a function of
$N_\tau$. The maximum value $g_{av}^{max}$ is independent of $N$ at
criticality, see Fig.~\ref{bindmim}. We extrapolate the thermodynamic
critical values by considering a collection of chains with sizes
ranging from $N=8$ to $N=32$.  For each $\Delta \tau_c \Gamma_c$ we
looked for the value of $\alpha$ such that the $N$-independent maximum
is reached. The phase diagram in Fig.~\ref{thermolimit} has been built
repeating these steps.  The error-bars estimate the dispersion on the
values of $g_{av}$ arising from the Monte Carlo method and the
disordered interactions after performing error propagation on
Eq.~(\ref{defbinder}).

In Fig~\ref{thermolimit} we see that the infinite chain has a phase
transition even at $\alpha=0$, in contrast to the small $N$ case. We
find that the phase transition curve may be accurately described by a
power law, and the size of the ordered phase increases with increasing
coupling to the environment.  For comparison we also show the
transition line for a non-random Ising chain with Ohmic
dissipation~\cite{Troyer} as determined by using the same procedure and
similar sizes as the ones we use in the disordered problem.

The critical values for $\Delta \tau_c \Gamma_c$ with $\alpha_c=0$,
which corresponds to switching off the bath, agree with well established
analytical results for both the random and non-random
cases~\cite{McCoy,Onsager}. Indeed, the critical values of the random
and non-random models in terms of our variables are $\Delta \tau_c
\Gamma_c=0.25$ and $\Delta \tau_c \Gamma_c=0.44$ respectively, which
agree with our numerical calculations~\cite{note}.

\subsubsection{The global linear susceptibility}
\label{sec:globallinear}

In a usual paramagnetic to ferromagnetic phase transition 
one can expect to identify the critical line 
by looking for the location of the divergence 
of the global linear susceptibility defined as  
\begin{equation}
\chi \equiv \sum_{i=1}^N \chi_i =
\sum_{i=1}^N \left[\left.
\frac{\partial \langle m_i\rangle }{\partial h_i} \right|_{h_i=0} 
\right]
=
\frac1{N_\tau}
\sum_{i=1}^N \left[ \langle m_i^2 \rangle - \langle m_i \rangle^2\right]
\;,
\label{eq:global-chi} 
\end{equation}
and
\begin{equation}
m_i = \sum_{t=1}^{N_\tau} s_i^t
\; .
\end{equation} 

\begin{figure}[t!]
\resizebox{12cm}{!}{\includegraphics*[1.5in,4.7in][7in,7.7in]{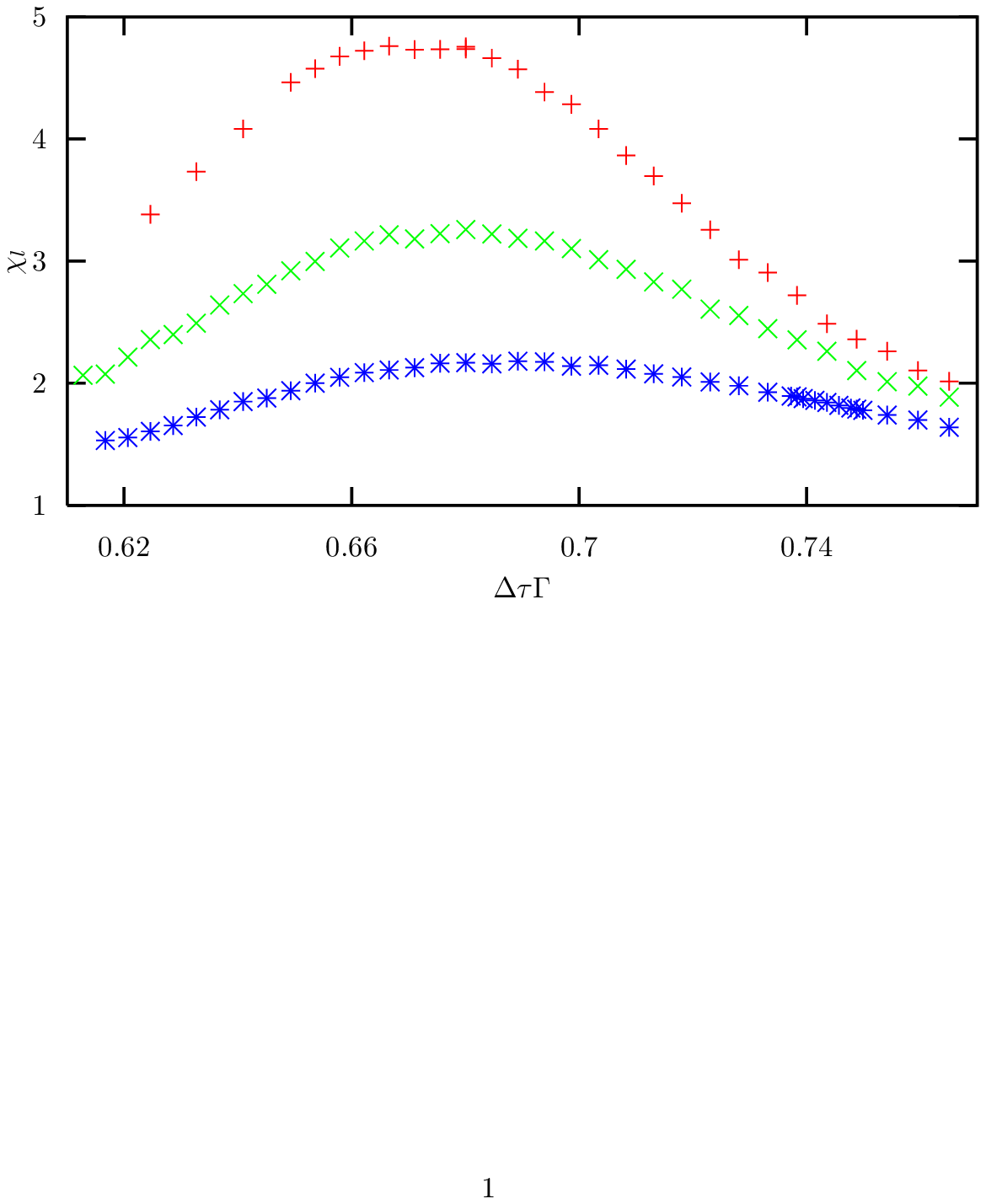}}
\hspace{-0.5cm}
\resizebox{12cm}{!}{\includegraphics*[1.5in,4.7in][7in,7.7in]{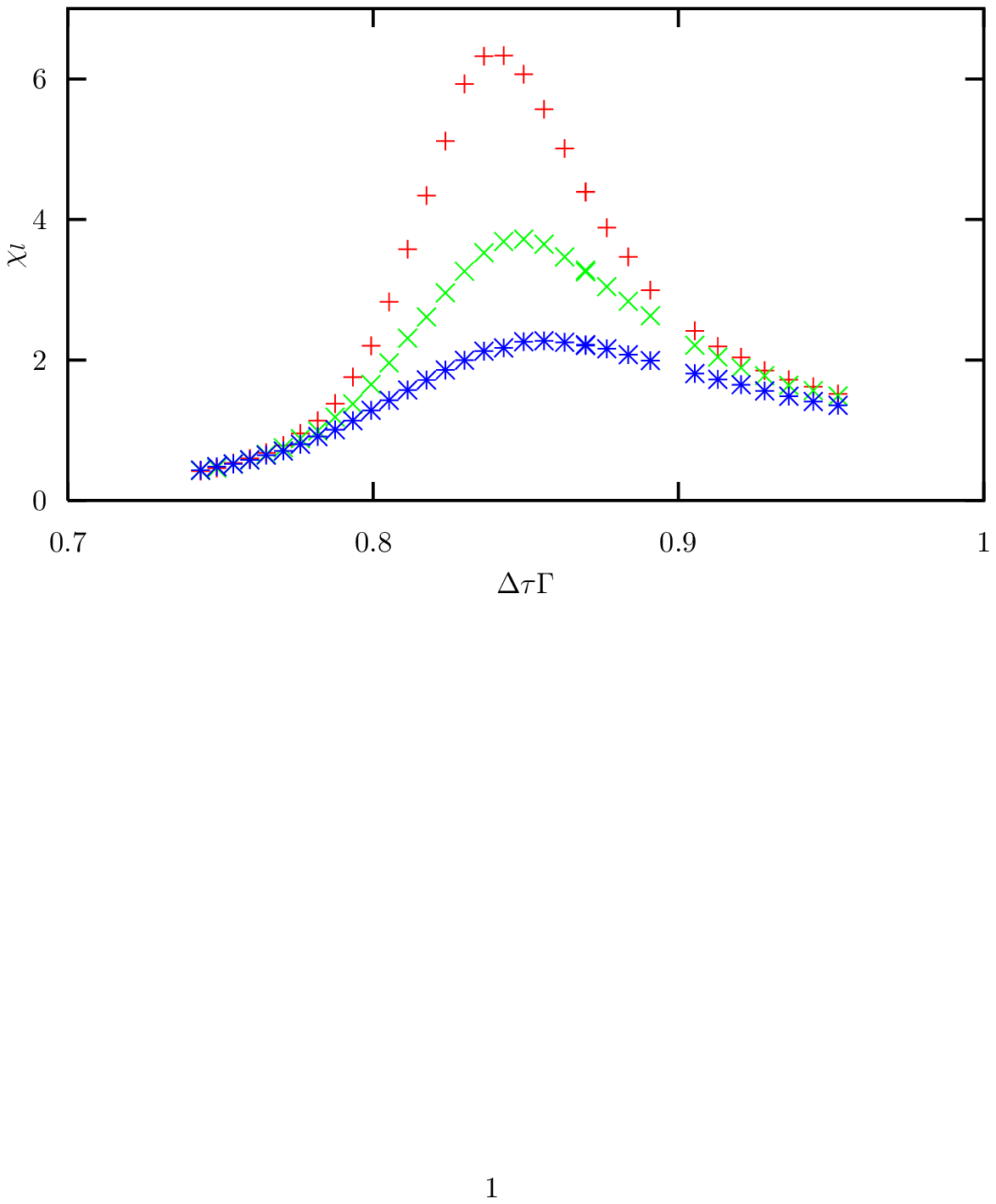}}
\caption{The average global linear susceptibility $\chi$ for
$\alpha_c=0.45$ as a function of $\Delta\tau \Gamma$. 
$N=4$ and $N_\tau=1 28,64,32$ from top to bottom.
Top panel: Random interactions. Bottom panel: pure ferromagnet.}
\label{sucu}
\end{figure}

With the aim of later studying the effect of dissipation on the
distribution of local susceptibility of the random chain, we attempted
to analyze the divergence of the global susceptibility in finite
clusters~\cite{Details0}.  Indeed, when $N<\infty$ we expect to find a
divergence of $\chi$ only on the phase transition, and not within the
paramagnetic phase as found for the isolated random infinite
chain~\cite{Fisher}.  For finite values of the number of
imaginary-time slices $N_\tau$ we found that the global linear
susceptibility at $\alpha$ fixed has a peak at value of
$\Delta\tau\Gamma$ that is quite higher than the one obtained from the
analysis of the Binder parameter in the limit $N_\tau\to\infty$. As
expected the height of the peak increases while its position moves
towards lower values of $\Delta\tau\Gamma$ when $N_\tau$
increases. For $N=4$, $N_\tau=128$ and $\alpha=0.45$ we find a deviation
of the order of $10\%$.  This indicates that the finite size
effects in $N_\tau$ are still very important. Note that this holds for
the ordered and disordered system as well.  These observations are
illustrated in Figs.~\ref{sucu}.

\begin{figure}[t]
\resizebox{12cm}{!}{\includegraphics*[1.5in,4.7in][7in,7.7in]
{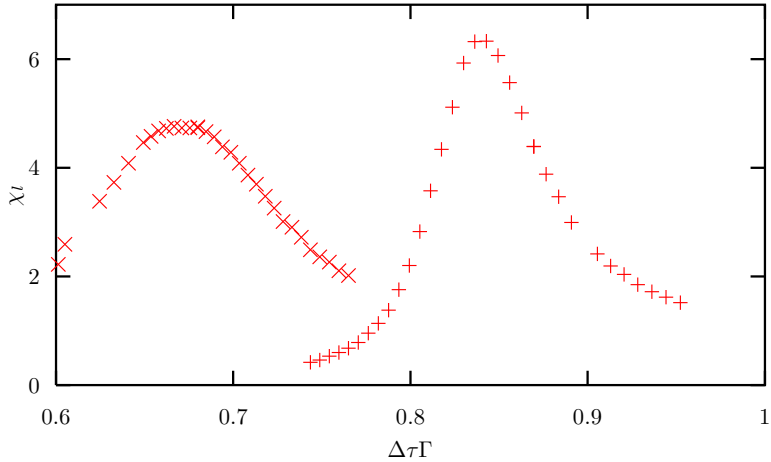}}
\caption{Comparison between the global linear 
susceptibility for
the model with random (left) and pure (right) systems 
with $N=4$, $N_\tau=128$ and $\alpha_c=0.45$.}
\label{suc_comparada}
\end{figure}

In Fig.~\ref{suc_comparada} we attempted to compare the form of the 
peak for similar ordered and disordered systems, 
{\it i.e.} with the same size and under the effect of the 
same external bath. We note that the peak is more pronounced
in the ordered case and the deviation between its location 
at finite $N_\tau$ and the asymptotic $N_\tau\to\infty$
one is less pronounced. More precisely, we found
\begin{eqnarray}
(\Delta\tau\Gamma)_c &=&
\left\{
\begin{array}{ccc}
\mbox{Binder} \; 
 & \mbox{Peak in} \; \chi \;\;\;\;\;\; &
\\
N_\tau\to\infty & N_\tau=128 \;\;\;\;\;\;& 
\\
0.77 & 0.84 & \mbox{ordered}
\\
0.55 & 0.66 & \mbox{disordered}
\end{array}
\right.
\end{eqnarray}

In conclusion, we found that the extrapolation of the finite $N_\tau$
susceptibility data to the thermodynamic limit is indeed very 
hard~\cite{N=1}. A detailed
analysis of the scaling behavior needs to be perform but it currently
goes beyond out computational power. This failure has to be kept in 
mind when trying to analyze the dynamic properties of the dissipative 
disordered Ising chain in the paramagnetic phase, see 
Sect.~\ref{sect:critical}.

\subsection{Critical scaling: conventional or activated?}
\label{sect:critical}

One of peculiarities of the critical behavior of the isolated 
quantum Ising chain with random interactions is that the 
critical scaling is activated instead of of conventional power 
law type. We wish to investigate whether a similar behavior 
persists when Ohmic dissipation is included.

\subsubsection{The Binder ratio}

The Binder ratio not only gives a criterion to find the phase transition
curve but also helps to derive scaling laws.  The study of the scaling
laws for the function $g_{av}$ can in principle provide an answer to
the question of whether the critical behavior is of conventional or
activated type.  Finite size scaling implies that dimensionless
quantities should scale as functions of $N/\xi$, and as either
$N_\tau/N^z$ in the case of conventional scaling or
$N_\tau/e^{N^{\overline z}}$ in the case of activated scaling.  At the
critical point, the correlation length in the spatial direction,
$\xi$, diverges. This suggests that the scaling variable should be
$N_\tau/N_\tau^{max}$ or $\ln \, N_\tau/ \ln \, N_\tau^{max}$ in each
case respectively, where $N_\tau^{max}$ is the value of $N_\tau$ that
maximizes $g_{av}$~\cite{Guoetal}.  
This way of analyzing the data has the advantage
of not needing to determine the critical exponents $z$ or
$\overline{z}$.

However the interpretation of the numeric results is delicate.  Even
for $\alpha=0$, where it is possible to show analytically the
activated scaling~\cite{Fisher}, it has been hard to distinguish
between both types of scaling by means of Monte Carlo
simulations~\cite{Rieger-young,Guoetal,Crisanti}. In particular, 
in $d=1$ the numerical data confirmed the analytic prediction
only when a mapping to fermions was used~\cite{Young-Rieger}.

\begin{figure}[t!]
\resizebox{12cm}{!}{\includegraphics*[1.5in,4.7in][7in,7.7in]{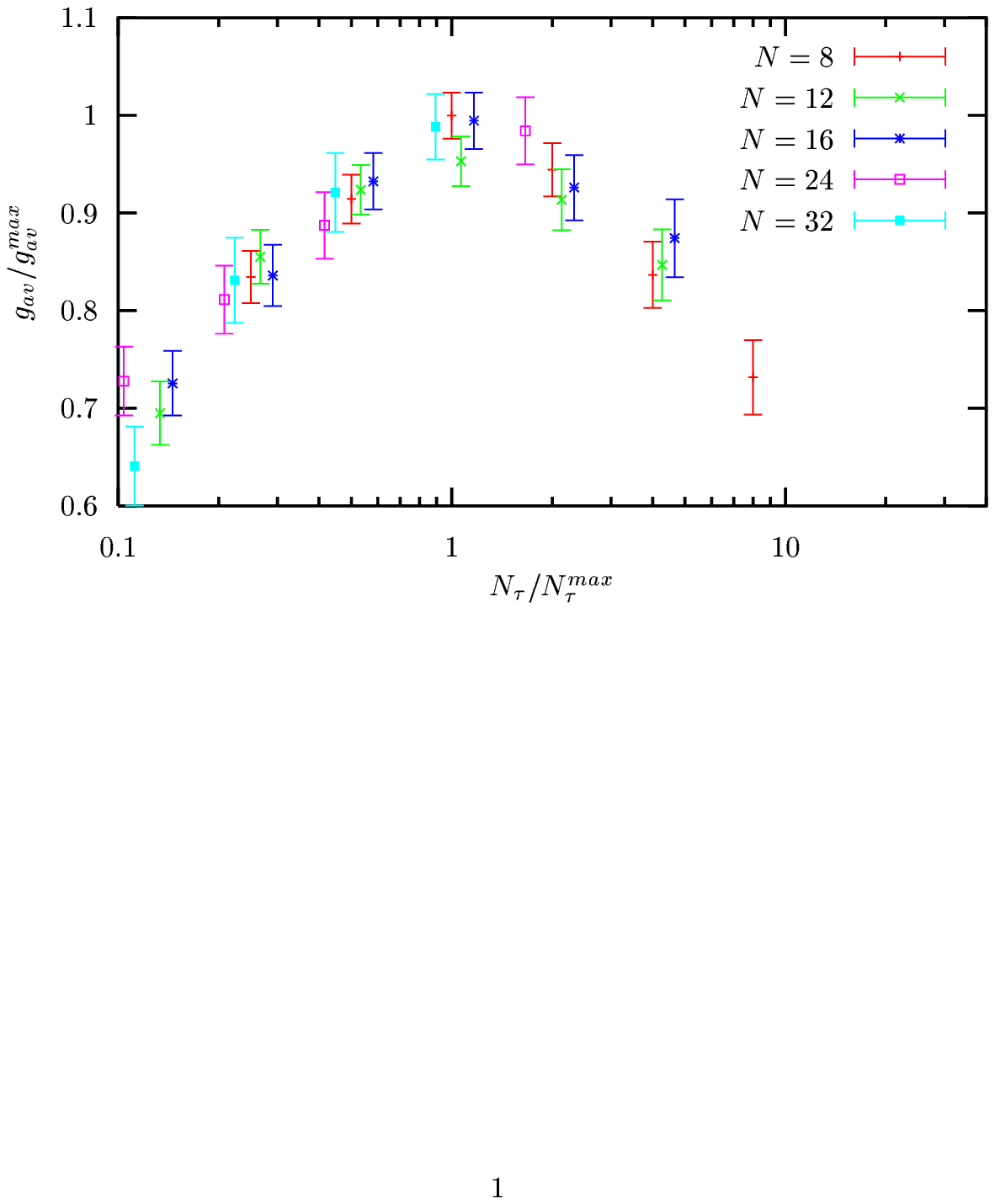}}
\caption{(Color on-line.)
Test of conventional  
scalings at $(\Delta \tau_c \Gamma_c=0.48, \alpha_c=0.25)$.
}
\label{gcomun0.4}
\resizebox{12cm}{!}{\includegraphics*[1.5in,4.7in][7in,7.7in]{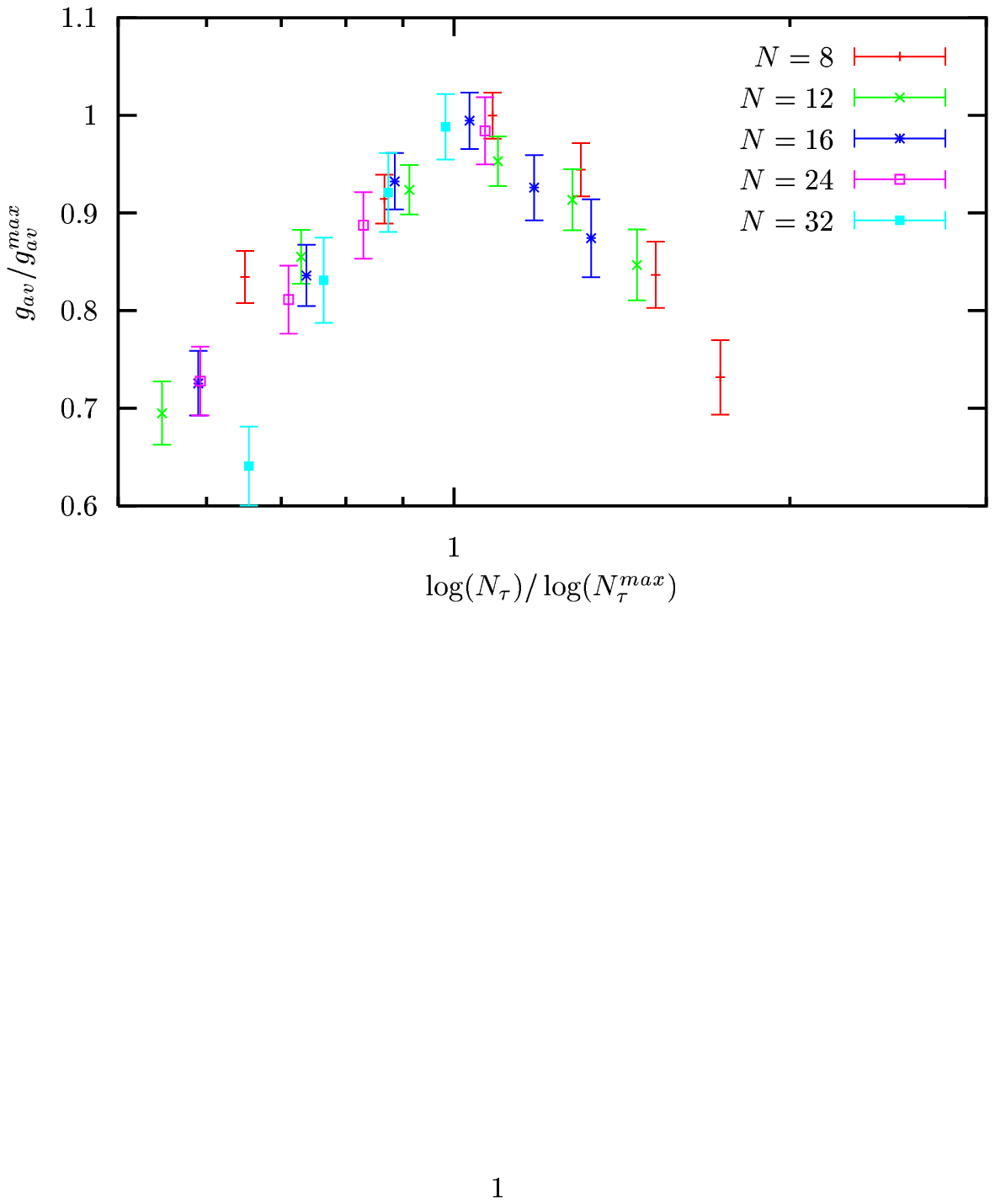}}
\caption{(Color on-line)
Test of activated 
scalings at $(\Delta \tau_c \Gamma_c=0.48, \alpha_c=0.25)$.
}
\label{gactivado0.4}
\end{figure}

In Fig~\ref{gcomun0.4} we show the scaling of the Binder parameter,
$g_{av}/g_{av}^{max}$, with $N/N_\tau^{max}$ (conventional) and in
Fig~\ref{gactivado0.4} with $\ln N/\ln N_\tau^{max}$ (activated). Both
scalings give almost the same critical values $(\Delta \tau_c
\Gamma_c,\alpha_c)$ in agreement with our previous simpler analysis
using the independence of the height of the maximum on $N$ at
criticality. Unfortunately, the two scaling plots are of quite similar
quality and with the system sizes considered here we cannot
distinguish between the two critical scalings.

\subsubsection{Distribution of local susceptibilities}
\label{sec:pdfs}

An alternative analysis of the critical scaling in the disordered 
models is based on the study of the probability 
distribution function (pdf) 
of local linear and non-linear susceptibilities and 
how they behave when approaching the critical line.

In the isolated random chain the pdf of local linear and non-linear
susceptibilities decay, for large values of the arguments, with a
power law that decreases when approaching the quantum critical
point. The inverse of this power is linked to the dynamic critical
exponent and its divergence on the transition implies activated
scaling.  At a finite distance from the critical point the decay
becomes sufficiently slow so as to lead to a divergent global
susceptibility. One may wonder whether this phenomenon also occurs in
the presence of dissipation.

\begin{figure}[p]
\begin{center}
\resizebox{10cm}{!}{\includegraphics*[1.5in,4.7in][7in,7.7in]{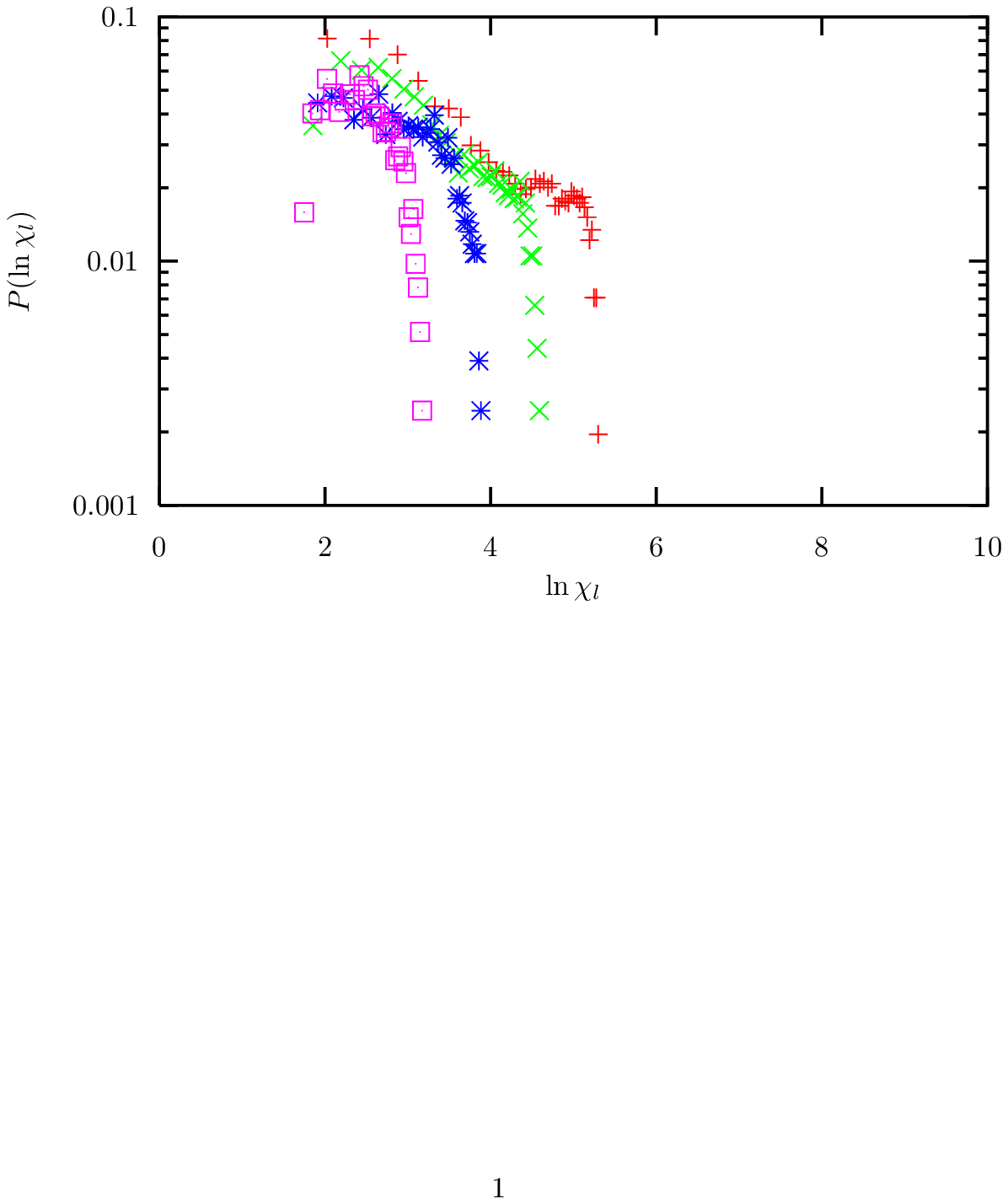}}
\end{center}
\caption{(Color on-line.) Probability distribution function of the 
local linear susceptibilities for a finite cluster with $N=4$ and
(from left to right)  
four values of the number of imaginary-time slices, $N_\tau=32, \, 64, \, 
128, \, 256$. $\alpha=0.45$ and $\Delta\tau\Gamma=0.66$.
}
\label{fig:problin}

\vspace{0.3cm}
\resizebox{0.5\textwidth}{!}{\includegraphics*[4.5cm,10.3cm][17cm,19cm]
{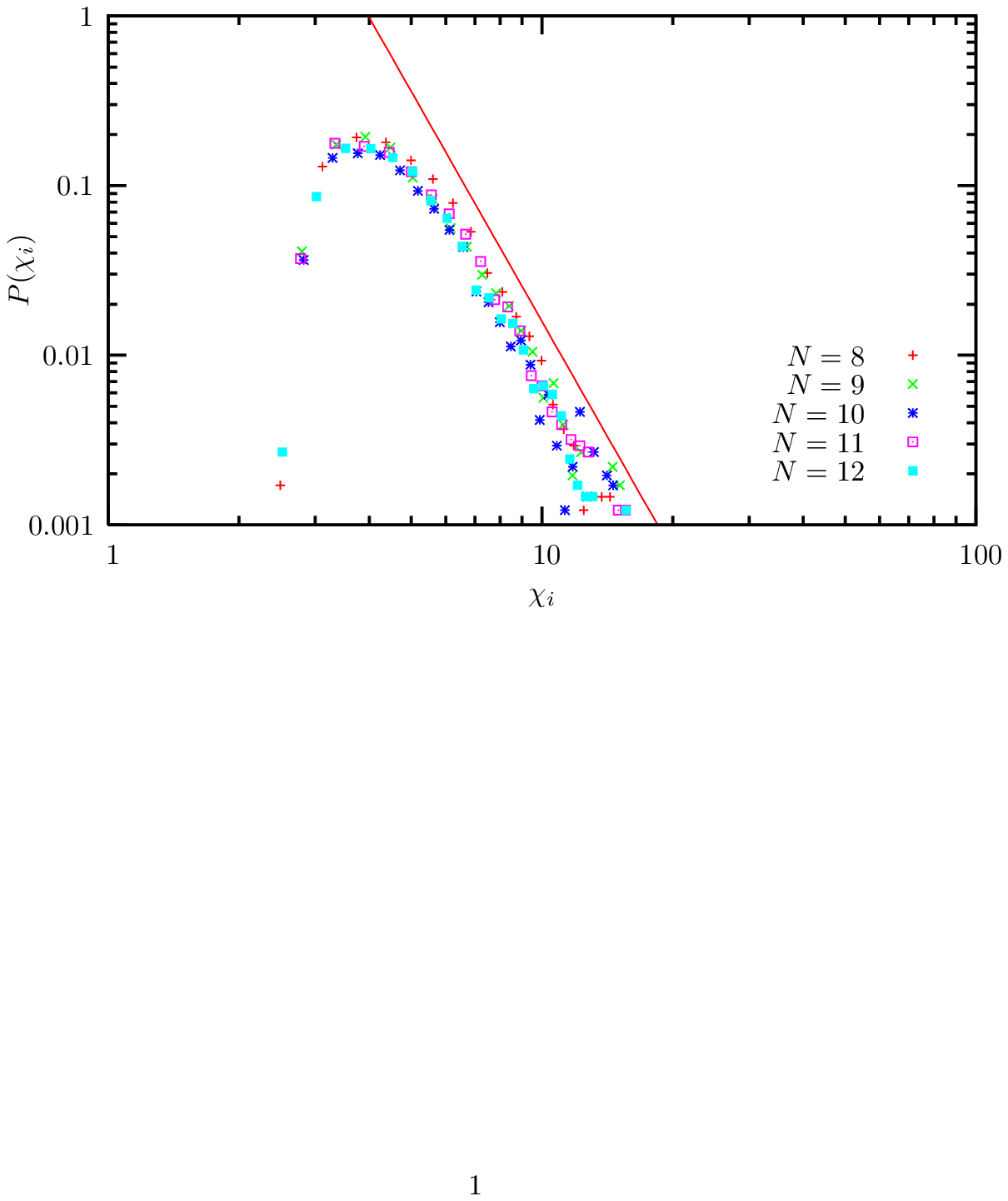}}
\resizebox{0.5\textwidth}{!}{\includegraphics*[4.5cm,10.3cm][17cm,19cm]
{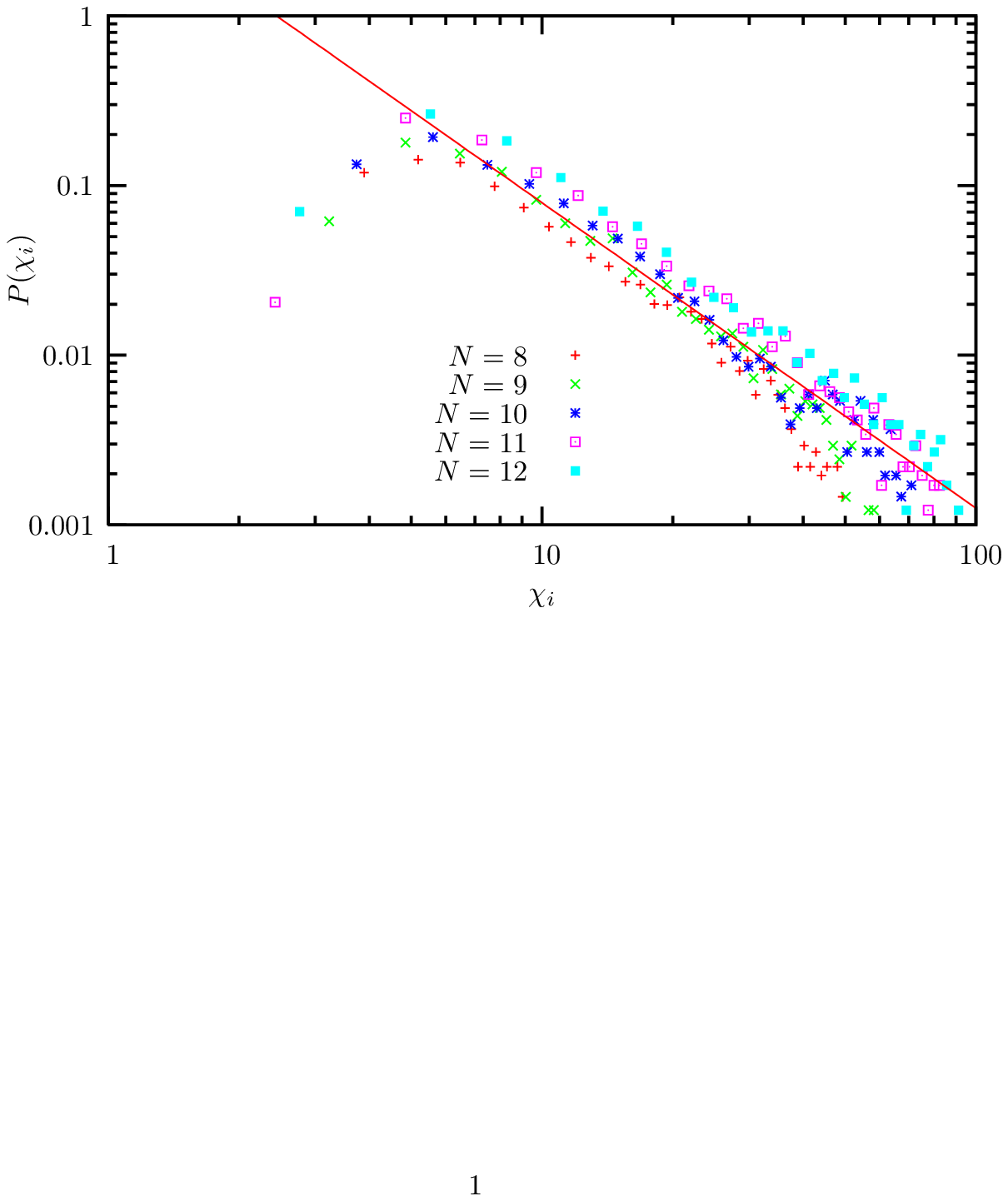}}
\resizebox{0.5\textwidth}{!}{\includegraphics*[4.5cm,10.3cm][17cm,19cm]
{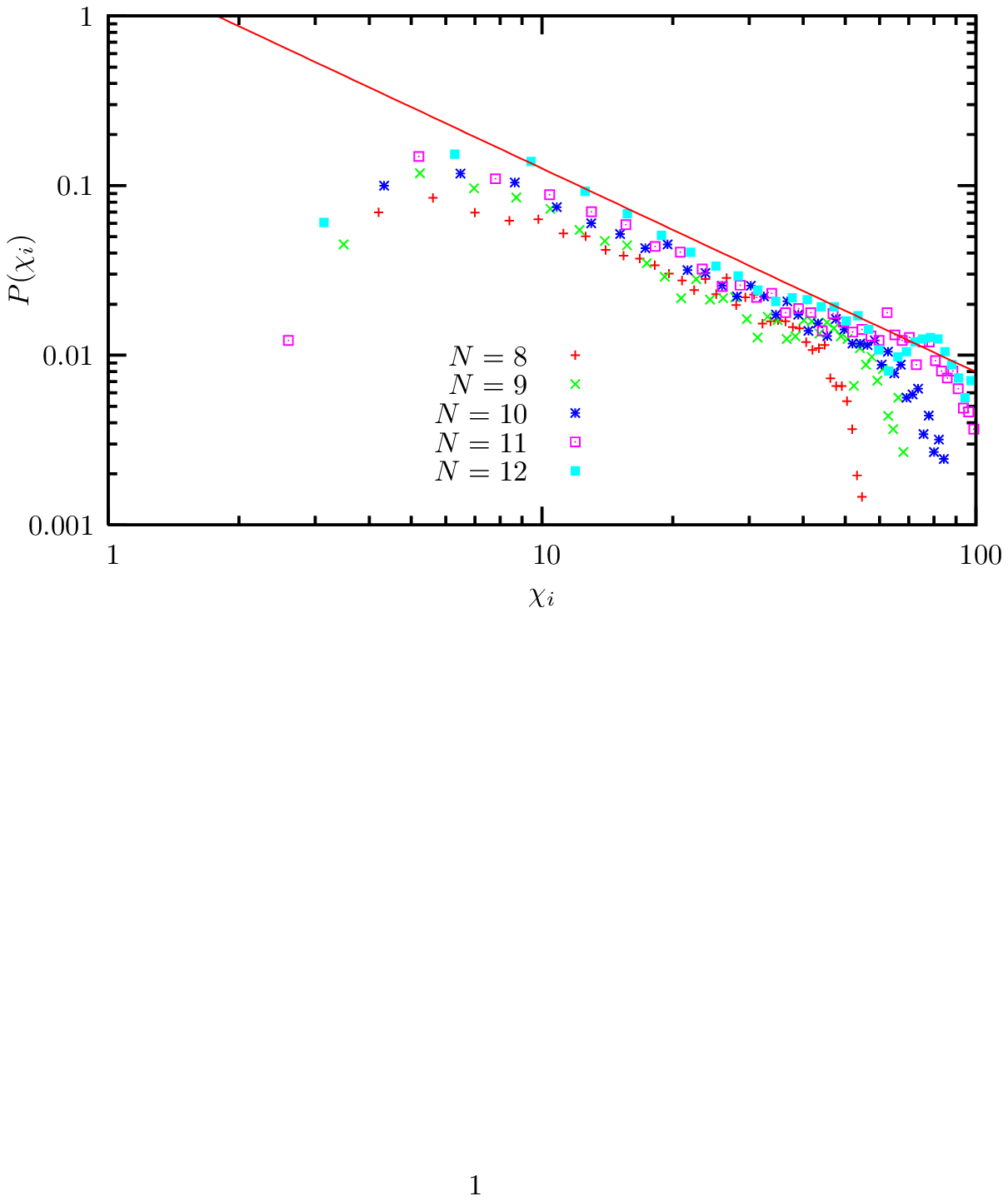}}
\resizebox{0.5\textwidth}{!}{\includegraphics*[4.5cm,10.3cm][17cm,19cm]
{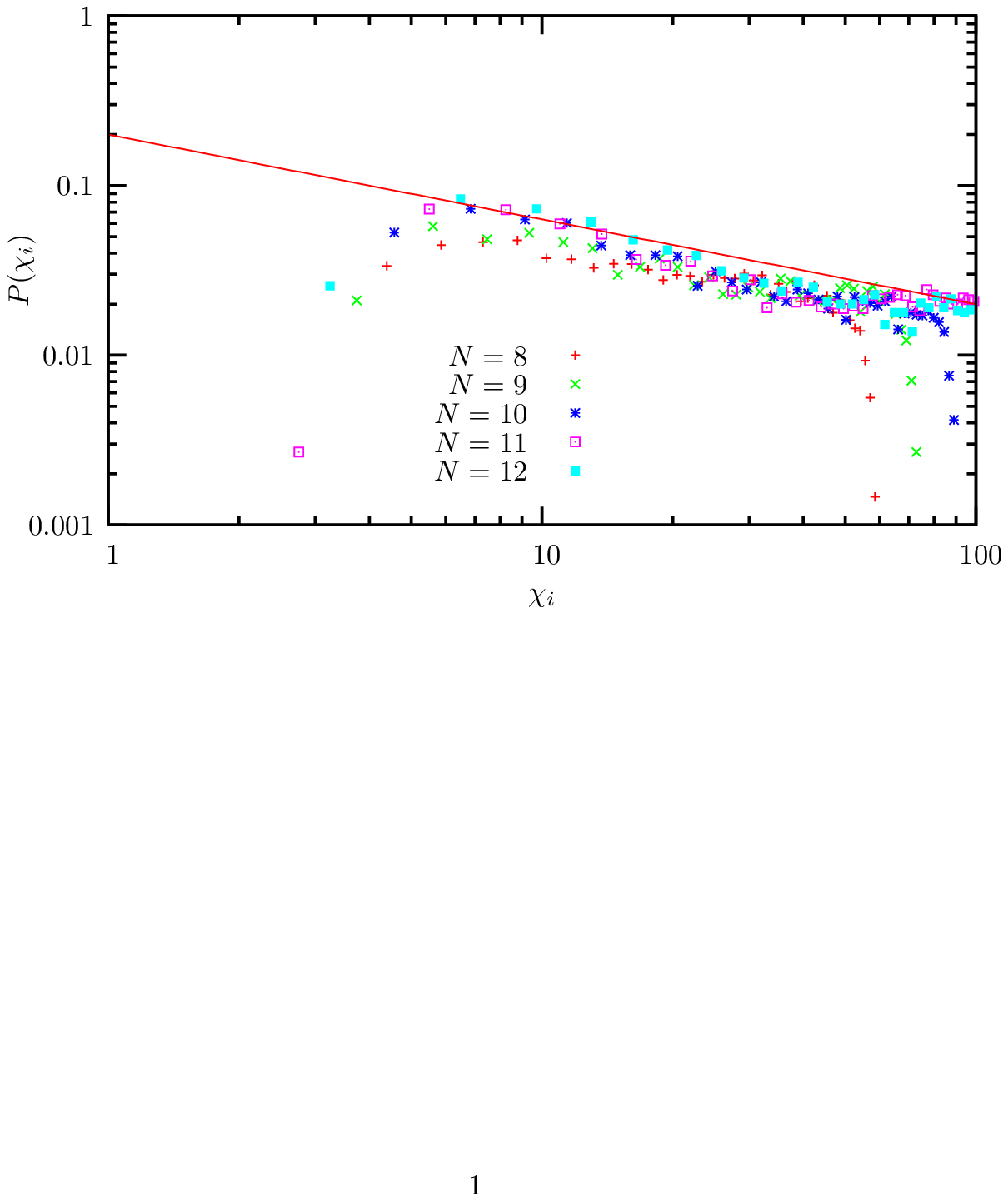}}
\caption{Probability distribution of the single site linear
susceptibility. $\Delta\tau\Gamma=0.48$ (top left), 
$0.43$  (top right), 
$0.41$  (bottom left), 
$0.39$ (bottom right). $\alpha=0.15$ in all cases.
The slope of the straight lines indicated in each panel 
are $b=4.55$ (top left), $1.82$ (top right), $1.20$ 
(bottom left), $0.5$ (bottom right).}
\label{fig:cuadrito}
\end{figure}

To try to give an answer to this question we started by 
studying the pdf of local linear susceptibilities  
of finite clusters ($N<\infty$).
In Fig.~\ref{fig:problin} we show $p(\ln\chi_i)$ for $N=4$ and four
values of the number of imaginary-time slices, $N_\tau=32,\, 64,\, 128, 
\, 256$, at $\alpha=0.45$ and $\Delta\tau\Gamma=0.66$. A finite 
value of $N_\tau$ gives a finite bound to the maximum possible 
$\chi_i$. This figure can be compared to the left-most 
curve in Fig.~20 in the first reference in 
\cite{Young-Rieger} that has been obtained for $N=4$ 
and effectively with $N_\tau\to\infty$. This figure shows that the 
coupling to the external bath has not modified the form of the 
pdf in an important way.

Finally, we analyzed the decay of the probability density of 
the local linear susceptibilities when approaching the critical line.
Each panel in Fig.~\ref{fig:cuadrito} displays these
pdfs for four values of $\Delta\tau\Gamma$ 
and $\alpha$ fixed. Each set of data, represented with different 
symbols, were obtained for different $N$ and choosing
$N_\tau$ in such a way that the maximum in the Binder 
parameter is reached. The values of $N$ are given in the captions.
The straight lines are guides to the eye to 
indicate a possible power-law decay and the values of this
power are given in the caption.

Even if it is clear from the 
figures that there is a  trend to a decreasing power law decay 
as the transition is approached, a 
functional relation between this power and the distance to the 
critical line is hard to establish beyond doubt from the 
pure numerical data. Analytic guidance, as was necessary to 
interpret correctly the numerical data for the isolated case, seems
to be necessary here too.

\section{Conclusions}
\label{sec:conclusions}

With a careful analysis of the Monte Carlo data we 
succeeded in determining the phase boundary between the  
paramagnetic and ferromagnetic phases in finite and infinite
rings of interacting quantum Ising spins in a transverse field and 
coupled to an external environment. While for finite number 
of spins there is no phase transition in the limit $\alpha\to 0$,
in the thermodynamic limit one recovers the critical
points of McCoy and Wu in the disordered case, and Onsager
in the ordered problem.

The method that we used allowed us to determine the Caldeira-Leggett
critical value $\alpha_c=1$ for $N=1$ with relatively little numerical
effort. We used a similar analysis to find the phase transition of
interacting finite clusters under the effect of an Ohmic bath.  A
similar analysis could be used to study the influence of other types
of baths (sub-Ohmic, super-Ohmic) for which less analytical results
are available.

The coupling to the bath enhances the extent of the ordered phase, as
found in mean-field spin-glasses~\cite{bagno1,bagno2} and in the
ordered ferromagnetic Ising chain~\cite{Troyer}.  The Griffiths phase
still exists as characterised by large distributions of the local
linear and non-linear susceptibilities (the latter are not shown in
the text). We expect to find similar results in higher finite
dimensions and when frustration is included ({\it i.e.} when negative
exchanges also exist).

Our results for the type of scaling characterizing the 
phase transition for the disordered and dissipative infinite chain
are not conclusive. Even if we have no evidence 
for the failure of the activated scaling of the isolated case 
when dissipation is included, we cannot exclude this 
possibility from our numerical data. It is important to 
notice that the activated scaling in the dissipation-less 
problem was found analytically 
and it was very hard to confirm numerically. It would be 
extremely interesting to extend Fisher's renormalization
group approach to the case of the random chain coupled to  
an environment to give a definitive answer to this question.

The systems we studied admit applications in a variety of fields. 
Quantum (spin) glass phases have been observed experimentally
in several compunds~\cite{qglass}; the role played by the coupling to the 
environment in their low-temperature anomalous properties has not 
been fully elucidated yet. Interacting spin systems are now
becoming popular as possible toy models for quantum computers, 
with the Ising spins representing the qubits. It is obviously 
very important in this context to establish the effect of the 
coupling to the bath. Another motivation for our study comes from 
the proposal in \cite{Antonio} to describe non-Fermi liquid behavior
in certain $f$-electron systems with a similar model (see 
also~\cite{Millis}). Finally, while classical phase transitions are
well understood by now, the same does not apply to quantum phase 
transitions at zero temperature~\cite{Sachdev,Senthil}. For all 
these reason we believe that this (and related ones) are interesting 
problems that deserve further study.

\vspace{1cm}
\noindent{\underline{Acknowledgements.}}  LFC is a member of the
Institut Universitaire de France and acknowledges financial support
from an Ecos-Sud travel grant, the ACI project ``Optimisation
algorithms and quantum disordered systems'' and the ICTP-Trieste, as
well as hospitality from the Universidad de Buenos Aires and the
Universidad Nacional de Mar del Plata, Argentina and ICTP-Trieste
where part of this work was prepared.  This research was supported in
part by SECYT PICS 03-11609 and PICS 03-05179 and UBACYT/x053. We 
especially thank Daniel Grempel for very useful discussions.


\begin{thebibliography}{99}
\bibitem{note} When $\alpha=0$, $B$ can be factorized in the effective
$2d$ classical action~(\ref{classicalA}) as an inverse effective
``temperature'' showing close correspondance to the McCoy-Wu random
Ising model~\cite{McCoy}. Thus, in the McCoy-Wu model with uniform
distribution of exchanges, the critical temperature, $T_c$, satisfies
$ 2/T_c + [\ln\tanh(J_i/T_c)] =0 $. Then, $B=1/T_c=0.70$ and through
Eq.~(\ref{defB}) we obtain $\Delta \tau_c \Gamma_c=0.25$. In
particular, for $J_i=1$ we recover the Onsager's result with
$B=1/T_c=0.44$ and through Eq.~(\ref{defB}) we obtain $\Delta \tau_c
\Gamma_c=0.44$.
\bibitem{Details0} We computed the global linear susceptibility shown
in Sect.~\ref{sec:globallinear} using the expression in the
right-hand-side of Eq.~(\ref{eq:global-chi}) and taking the
magnetization with an absolute value in order to take into account the
ergodicity breaking in the ferromagnetic phase~\cite{Landau}. This
introduces a systematic shift in the peak of the susceptibility due to
the fact that $\langle m_i\rangle$ is not strictly zero on
paramagnetic side, as it should.
\bibitem{N=1} Indeed, the susceptibility of 
the dissipative single quantum Ising spin for finite $N_\tau$ 
also has a peak at a finite value of $\Delta\tau\Gamma$ that 
moves extremely slowly towards smaller values when $N_\tau$ increases. 
However, the values of $N_\tau$ that one can easily reach numerically
are not sufficient to safely extrapolate to zero. Another reason for 
this defect is that the way in which we computate $\chi$ (explained
in \cite{Details0}) overestimates its value in the paramagnetic 
phase.
\bibitem{Details1} Since we are here interested in the paramagnetic
phase only we computed the local linear susceptibility shown in
Sect.~\ref{sec:pdfs} using $\chi_i = N_\tau^{-1} \langle m_i^2\rangle$
(and assuming $\langle m_i \rangle=0$).
\end{thebibliography}
\end{document}